\documentclass[apj]{emulateapj}
\def\simgr{\,\hbox{\hbox{$ > $}\kern -0.8em \lower 1.0ex\hbox{$\sim$}}\,}
\def\simle{\,\hbox{\hbox{$ < $}\kern -0.8em \lower 1.0ex\hbox{$\sim$}}\,}

\newcommand{\rsolar}{\rm R_{\odot}}
\newcommand{\msolar}{$\rm M_{\odot}$}
\newcommand{\Ms}{${\rm M}_{\rm s}$}
\newcommand{\Rp}{${\rm R}_{\rm p}$}

\newcommand{\phx}{{\tt PHOENIX}}

\newcommand{\ang}{\hbox{\AA}}
\newcommand{\taustd}{\tau_{1.2}}
\usepackage{rotating,psfig,natbib}

\shortauthors{Barman et al.}
\shorttitle{Irradiated M dwarfs}

\slugcomment{Accepted for publication in the ApJ: June 24, 20004}
\begin{document}
\bibliographystyle{apj}

\title{Model Atmospheres for Irradiated Stars in pre-Cataclysmic Variables}

 \author{Travis S. Barman}
 \affil{Department of Physics, Wichita State University, Wichita, KS 67260-0032\\
        Email: {\tt travis.barman@wichita.edu}}
 \author{Peter H. Hauschildt}
 \affil{Hamburger Sternwarte, Gojenbergsweg 112, 21029 Hamburg, Germany\\
        Email: {\tt yeti@hs.uni-hamburg.de}}
 \author{France Allard}
 \affil{C.R.A.L, Ecole Normale Superieure, 69364 Lyon Cedex 7, France\\
        EMail: {\tt fallard@ens-lyon.fr}}

\begin{abstract} 

Model atmospheres have been computed for M dwarfs that are strongly irradiated
by nearby hot companions.  A variety of primary and secondary spectral types
are explored in addition to models specific to four known systems: GD 245, NN
Ser, AA Dor, and UU Sge.  This work demonstrates that a dramatic temperature
inversion is possible on at least one hemisphere of an irradiated M dwarf and the
emergent spectrum will be significantly different from an isolated M dwarf or a
black body flux distribution.  For the first time, synthetic spectra suitable
for direct comparison to high-resolution observations of irradiated M dwarfs in
non-mass transferring post-common envelope binaries are presented.  The effects
of departures from local thermodynamic equilibrium on the Balmer line profiles
are also discussed.

\end{abstract}


\section{Introduction}
Synthetic spectra from model atmospheres are frequently used in the analysis of
observed spectroscopic and photometric data. For the most part, the models are
sufficiently detailed to test the current theoretical understanding of stellar
and sub-stellar mass objects at various stages in their evolution.  However,
the majority of model atmospheres are intended for comparisons to isolated
stars and are not appropriate for many short period binaries.  A number of
binary systems have orbital separations small enough so that one of the binary
members is significantly heated by its companion.  In order for synthetic
spectra to be useful in such cases, the standard ``isolated" modeling approach
must be replaced by one that includes the effects of irradiation.  

Post-common envelope binaries (PCEBs) are examples of systems where large
effects due to irradiation have been observed (e.g. Z Cha, Wade \& Horne
(1988); GD 245, Schmidt et al. (1995); GD 444, Marsh \& Duck (1996)).  Many
PCEBs are cataclysmic variables (CVs) with secondaries that, in addition to
being irradiated, have over filled their Roche lobe and transfer mass to the
primary via accretion streams or a disk.  CVs, while very interesting, are not
necessarily the best choice for studying the effects of irradiation on cool
stars.  PCEBs that do not have on-going mass exchange offer several advantages.
The absence of accretion ensures that the primary is the only major source of
external heating for the secondary.  CVs have additional sources of radiation
energy (e.g., from the accretion disk) that are complicated to model and could
lead to shadows or bright spots on the secondary.  An additional advantage of
PCEBs without mass exchange is that the secondaries have not over filled their
Roche lobe and are therefore more spherical.  Most non-mass transferring PCEBs
are labeled pre-CVs and often contain a main sequence (MS) star in close orbit
around a much hotter white dwarf (WD) or sub-dwarf (sdOB).  The current study
will be restricted to pre-CVs with orbital periods less than 16 days or orbital
separations less than about 3$\rsolar$ (i.e., those that could possibly become
a CV within a Hubble time; \cite{hillwig00}). For an excellent review of
detached binaries containing WD primaries, see \cite[]{marsh00}.

The average effective temperature of the primary (the WD or sdOB) in pre-CVs is
about 50,000K with a few reaching 100,000K.  At just a few $\rsolar$ away, the
much cooler secondary star is significantly heated by the primary's radiation.
Therefore, the secondary's irradiated atmosphere is regulated by both {\it
extrinsic} radiation from the primary and {\it intrinsic} energy supplied by
internal nuclear reactions (or, in the case of a brown dwarf secondary,
leftover gravitational energy from formation).  Depending on the temperatures,
pressures, and chemical composition of the secondary's atmosphere, a fraction
of the extrinsic radiation will be reflected via scattering by grains,
molecules or electrons (or a combination of these) while the remaining is
absorbed and re-radiated.  Furthermore, the two sources of energy may have
spectral energy distributions (SEDs) that peak at very different wavelengths
which implies that a broader category of opacity sources (spanning the EUV to
the far IR) become important in shaping the seconday's atmospheric structure
and emergent spectrum.  The situation is also complicated by the fact that only
one hemisphere of the irradiated companion is heated at any given time.  

Only a few theoretical studies applicable to the atmospheres and spectra of
pre-CVs have been published.  One of the more recent works investigated the
effects of irradiation in the eclipsing binary BE Ursae Majoris using {\tt
CLOUDY} \cite[]{Ferguson1994}.  A narrow range of M dwarfs ($3400\rm{K} < T_{\rm
eff} < 3800\rm K$) located near a 10,000K black body have also been modeled
\cite[here after BS93]{Brett93}.  A few earlier papers provided critical
insight into the problem of irradiation but did not consider situations
relevant for pre-CVs \cite[]{VazNordlund1985, NordlundVaz1990}.  In addition,
the necessary opacity data and spectral line databases needed to compute
realistic synthetic spectra were not available at the time.  Consequently, no
synthetic spectra have been published that are detailed enough to provide
useful comparisons to high-resolution, phase-resolved, observations of pre-CVs.
Furthermore, most light curve models continue to use black body SEDs or, at
best, non-irradiated atmosphere models.

In the following sections, models for a variety of irradiated pre-CV
secondaries will be presented.  The changes in the atmospheric structure, the
chemical composition, and the spectra will be described.  Detailed,
line-blanketed, synthetic spectra will also be presented for a few specific
pre-CV systems. 

\section{Model Construction}\label{modcon}

All calculations presented here were produced using the \phx\ model
atmosphere code \cite[]{jcam}.  \phx\  includes an equation of state and
radiative transfer solver that is suitable for modeling a broad range of both
hot \cite[]{jasonwinds} and cool \cite[]{AMES-2001} objects across the H-R
diagram.  This flexibility makes \phx\ well suited for modeling systems that
contain two extremely different objects (e.g. M dwarfs and WDs).

\begin{figure}
\plotone{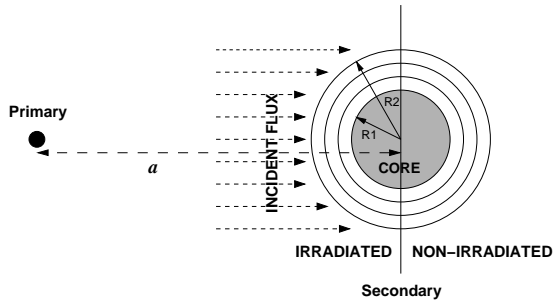}
\caption{
An illustration of the irradiation model. The primary (left) is treated as a
point source at distance $a$ from the center of the secondary (right).  The
secondary is divided into two hemispheres: irradiated and non-irradiated.  The
atmosphere on either hemisphere is modeled as a collection of concentric shells
with the height of the atmosphere equal to R2 - R1.
\label{irrad1}}
\end{figure}

The secondary may be conceptually divided into two distinct hemispheres;
irradiated and non-irradiated (see Figure \ref{irrad1}).  The irradiated
hemisphere receives incoming radiation over all angles between 0$^{\circ}$ and
90$^{\circ}$ with respect to the surface normal.  The effects of heating by the
primary will vary across the surface with decreased heating near the boundary
between the two hemispheres.  In order to simplify the simulations, each
hemisphere was modeled separately using a 1-D, spherically symmetric,
atmosphere designed to represent the {\em average} thermal and spectroscopic
properties.  Since pre-CVs are non-mass transferring systems, the secondary
does {\em not} overflow its Roche lobe and, hence, should be reasonably well
approximated by a sphere.  Consequently, the same mass and effective gravity
(set at a reference optical depth) was used for both hemisphere models.
However, despite having the same bulk properties, maintaining a consistent
temperature structure and entropy ($S$) at the bottom of both atmosphere models
for the irradiated and non-irradiated hemispheres is not guaranteed.
Differences between the temperature structures at the bottom of each model can
exist because the effects of horizontal energy flow have not been included and
the two hemispheres have been essentially decoupled.  While horizontal energy
flow is very likely to be present, including this in the simulations would
require the solution of the full, multi-D, radiation--hydrodynamical problem
that is currently beyond the scope of this paper.  The standard approach for
dealing with horizontal energy flow is to adjust the effective temperature of
one model so the entropy at the deepest layer is the same for both hemispheres
(Vaz \& Nordlund 1985\nocite{VazNordlund1985}; BS93\nocite{Brett93}).  The
details of this problem and how it is handled are discussed in a subsequent
section.  

All models were computed assuming hydrostatic and radiative-convective
equilibrium.  Convection was included using the standard mixing length theory
with the mixing length parameter set equal to two.  The atmospheres were
divided into 100 concentric shells and the spherically symmetric radiative
transfer equation was solved while explicitly including the incident, frequency
dependent, radiation field from the primary.  Spherical symmetry, as opposed to
the more traditional plane parallel geometry, was chosen because it provides a
more realistic description of the incoming radiation at grazing angles that
should not penetrate too deeply into the atmosphere. At some angles, the
radiation can pass completely through the top of the atmosphere.  The details
of the radiative transfer solution may be found in \cite{jcam} (and refs.
therein) and details of the irradiation are found in \cite{Barman01, Barman02};
however, for clarity, a few of the important points are repeated.

\begin{figure}
\plotone{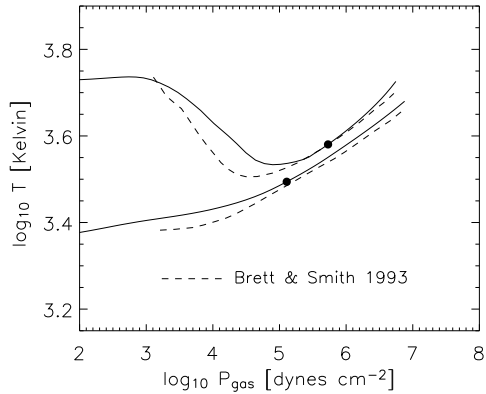}
\caption{
Temperature versus pressure for irradiated (top) and non-irradiated (bottom)
atmosphere models.  Solid lines are the results of \phx\ simulations while
dashed lines are taken from Fig. 3 of \cite{Brett93}.  Filled circles indicate
the location of the radiative-convective boundary in the \phx\ models.
The differences between the two sets of calculation are most likely due to
the different treatments of the equation of state and opacities.
\label{brett1}}
\end{figure}

A secondary with radius $R_{\rm s}$ intercepts an incident luminosity from a
primary of radius $R_{\rm p}$  at distance $d$ with flux $F_{\rm p}$ given by
$\pi R_{\rm s}^2\left(\frac{R_{\rm p}}{d}\right)^2 F_{\rm p}$.  This absorbed
energy must be distributed over and re-radiated by some fraction of the full
secondary surface area (4$\pi R_{\rm s}^2$).   The three most commonly explored
physical scenarios are re-radiation by the full surface area, by the irradiated
hemisphere only, or by a single point.  Since the models must conserve energy,
the incoming luminosity must balance the out-going luminosity of re-radiated
extrinsic energy.  In which case, $\pi  R_{\rm s}^2\left(\frac{ R_{\rm
p}}{d}\right)^2  F_{\rm p} = Area \times  F_{\rm out}$.  If one further assumes
that $F_{\rm inc} = F_{\rm out}$, the monochromatic incident fluxes can be
expressed as,
\begin{equation}
 F_{\rm inc,\lambda} =
\alpha \left(\frac{R_{\rm p}}{d}\right)^2  F_{\rm p,\lambda},
\end{equation}
where $F_{\rm p,\lambda}$ are the monochromatic fluxes from the primary
surface, and $d$ is the distance from the primary surface to the secondary
surface (note, $d = a - R_{\rm p} - R_{\rm s}$, where $a$ is the orbital
separation).  The parameter $\alpha$ sets the redistribution of the absorbed
incident energy over the secondary's surface.  All incident energy being
absorbed and re-radiated by the heated hemisphere corresponds to $\alpha =
0.5$.  Re-radiation by the entire secondary surface corresponds to $\alpha =
0.25$ and $\alpha = 1.0$ is for re-radiation by a single point on the surface.
For a more detailed description of the energy balance in an irradiated binary
companion and the development of a similar $\alpha$ parameter, see
\cite{paczynski80}. Unless otherwise stated, the models presented below were
calculated with $\alpha = 0.5$. This choice is motivated by the fact that many
short period binaries have tidally locked secondaries.

In all cases, the incident radiation was approximated by an isotropic radiation
field consisting of 64 in-coming intensities (corresponding to 64 incident
angles) at the top most layer ($\taustd = 0$, where $\taustd$ refers to a
reference optical depth measured at 1.2\micron) of the model.  Isotropic
illumination (instead of heating at a fixed angle) was chosen because it better
approximates the situation illustrated in Fig.  \ref{irrad1}.  While this is
clearly a simplification of the true incident radiation, isotropy roughly
accounts for the fact that the heated hemisphere receives radiation from the
primary at all angles between 0 and 90 degrees (with respect to the surface
normal).  Even when the incident radiation is assumed to be isotropic, there
are still benefits to using spherical geometry.  Spherical geometry, combined
with isotropic irradiation, accounts for the preferential heating of the upper
atmospheric layers compared to the deeper layers for regions near the boundary
between the two hemispheres.  Across the heated hemisphere, the upper layers
always receive extrinsic radiation while the deeper layers experience
considerably less heating near the boundary compared to the substellar point.
By virtue of the translational symmetry inherent in plane parallel models, {\em
all} inward directed intensities are propagating toward the center of the
object. Thus, for some situations, plane parallel geometry could overestimate
the heating of deeper atmospheric layers and underestimate the heating of upper
layers.  The differences between plane parallel and spherical models are
maximized when the atmospheric extension is large compared to the mean free
path of a photon (e.g., in low gravity objects). For a spherical model, a large
extension has the effect of decreasing the range of angles over which incident
intensities actually propagate toward the bottom of the photosphere.  In the
future, the heated hemisphere will be modeled as a collection of small annular
regions where the extrinsic radiation is incident along a single angle (or a
narrow range).  In this case, the advantages of spherical geometry over plane
parallel will be even more significant.

Since pre-CV primaries have spectra that are very different from black body
SEDs, the incident intensities were taken from synthetic WD spectra (also
calculated with \phx; Barman et al. 2000)\nocite{barman00}.  In this way, a
more realistic, wavelength dependent spectrum was used for the incoming
radiation.  Unless otherwise stated, the WD models have $\log(g) = 8.0$ (cgs
units), 1 \msolar, and $10^{-2}$ solar metal abundances.  When modeling a
secondary for a specific pre-CV, a new model atmosphere and synthetic spectrum
were used that closely match the primary spectral type and, if available, the
observed fluxes.

The model calculations for both the primary and the secondary included more
than 50,000 wavelength points between 10\AA\ and 1000\micron\ with a resolution
of about 1\AA\ from the UV to the near-IR (this resolution may be increased as
necessary).  The equation of state and opacity setup was similar to that used
in the NextGen model atmosphere grid of \cite{Nextgen99} but has since been
updated to include more recent chemical and opacity data \cite[]{AMES-2001}.
Local thermodynamic equilibrium (LTE) was assumed for the majority of the
calculations; however, the effects of non-LTE are explored for the hydrogen
atom.  All models for the secondaries used a solar composition that included 40
of the most important atomic elements from H to La (atomic numbers 1 through
57) as well as their important ions.  The atomic data for the energy levels and
bound-bound transitions are from \cite{cdrom22} and \cite{cdrom23}.  The
molecular opacities include H$_{\rm 2}$O \cite[]{ames-water-new}, TiO
\cite[]{ames-tio}, VO (linelist provided by R. Freedman, 2001, private
communication) all diatomics from \cite{cdrom15}, and all lines from the HITRAN
and GEISA databases \cite[]{hitran92, geisa}.  Collision induced absorption
(CIA) opacities are also included for H$_{\rm 2}$, N$_{\rm 2}$, Ar, CH$_{\rm
4}$, and CO$_{\rm 2}$ according to \cite[][and references therein]{co2co2}.
The total number of atomic and molecular lines currently available in \phx\
(version 13) is $\sim 700$ million.  Scattering (Thomson, Rayleigh, and Mie) was
also included and assumed to be isotropic.

The \phx\ model atmosphere code uses the effective temperature as an input
parameter that specifies (through the Stefan-Boltzmann formula) the net flux
(or luminosity) that the model will have.  To avoid confusion between the
effective temperature of irradiated and non-irradiated models, the variable
$T_{\rm eff}$ will be used only to refer to the effective temperature of
non-irradiated models.  In the case of irradiated models, $T_{\rm int}$ will be
used to describe the {\em intrinsic} effective temperature; i.e. the effective
temperature the model would have if irradiation was not present.  

\begin{figure}
\plotone{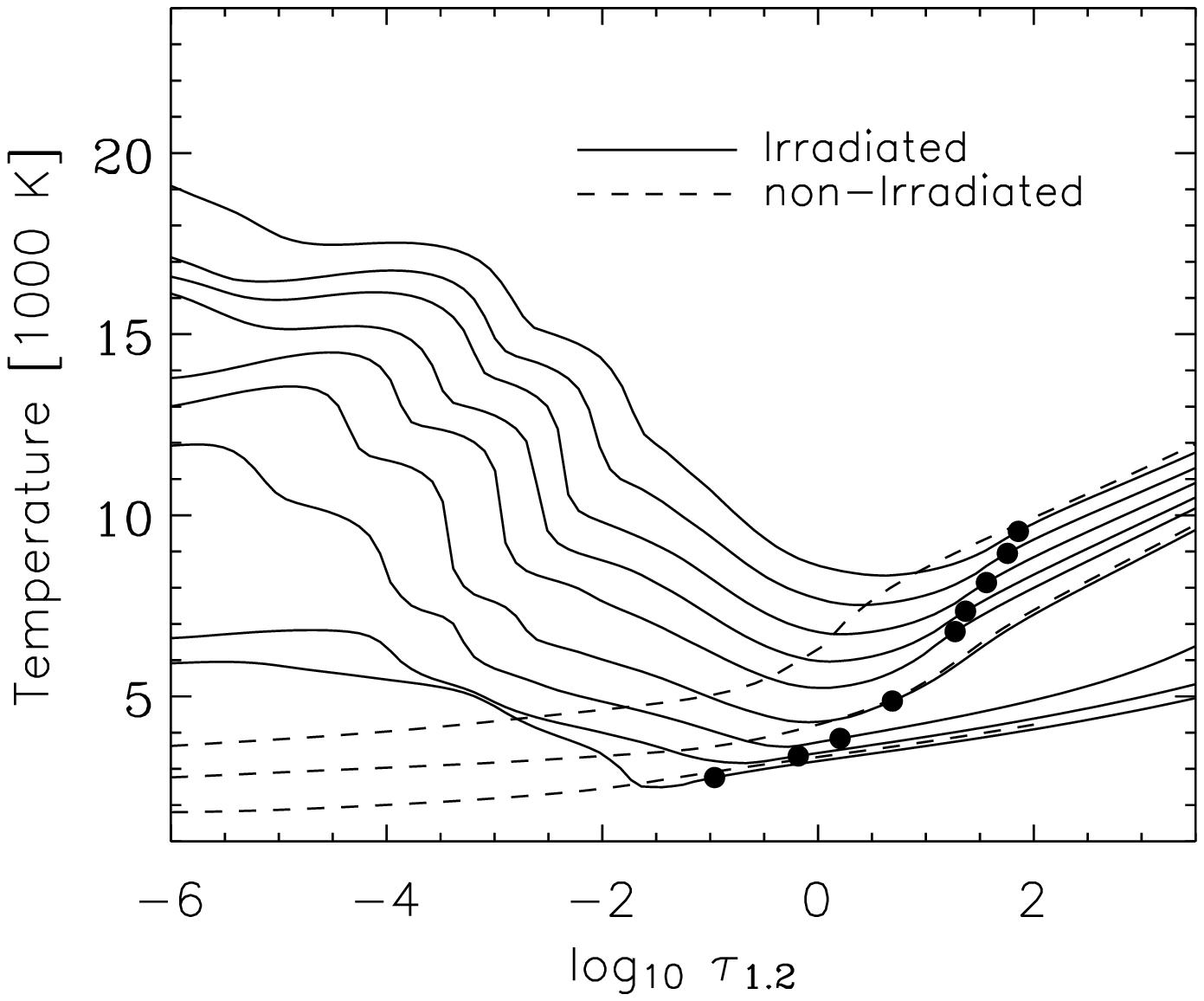}
\caption{
Temperature as a function of optical depth at 1.2\micron\ for irradiated (solid
lines) and non-irradiated (dashed lines) atmospheres.  The irradiated
atmospheres have $T_{\rm p} = $ 20,000K to 100,000K (from top to bottom) in steps
of 10,000K.  For all irradiated models, the orbital separation is 1 $\rsolar$
and $T_{\rm int} = 3000\rm K$. The non-irradiated models have $T_{\rm eff} = $ 5600K,
4000K, and 3000K (from top to bottom). Filled circles indicate the location of
the radiative-convective boundary.
\label{strucs1}}
\end{figure}

\begin{figure*}
\plotone{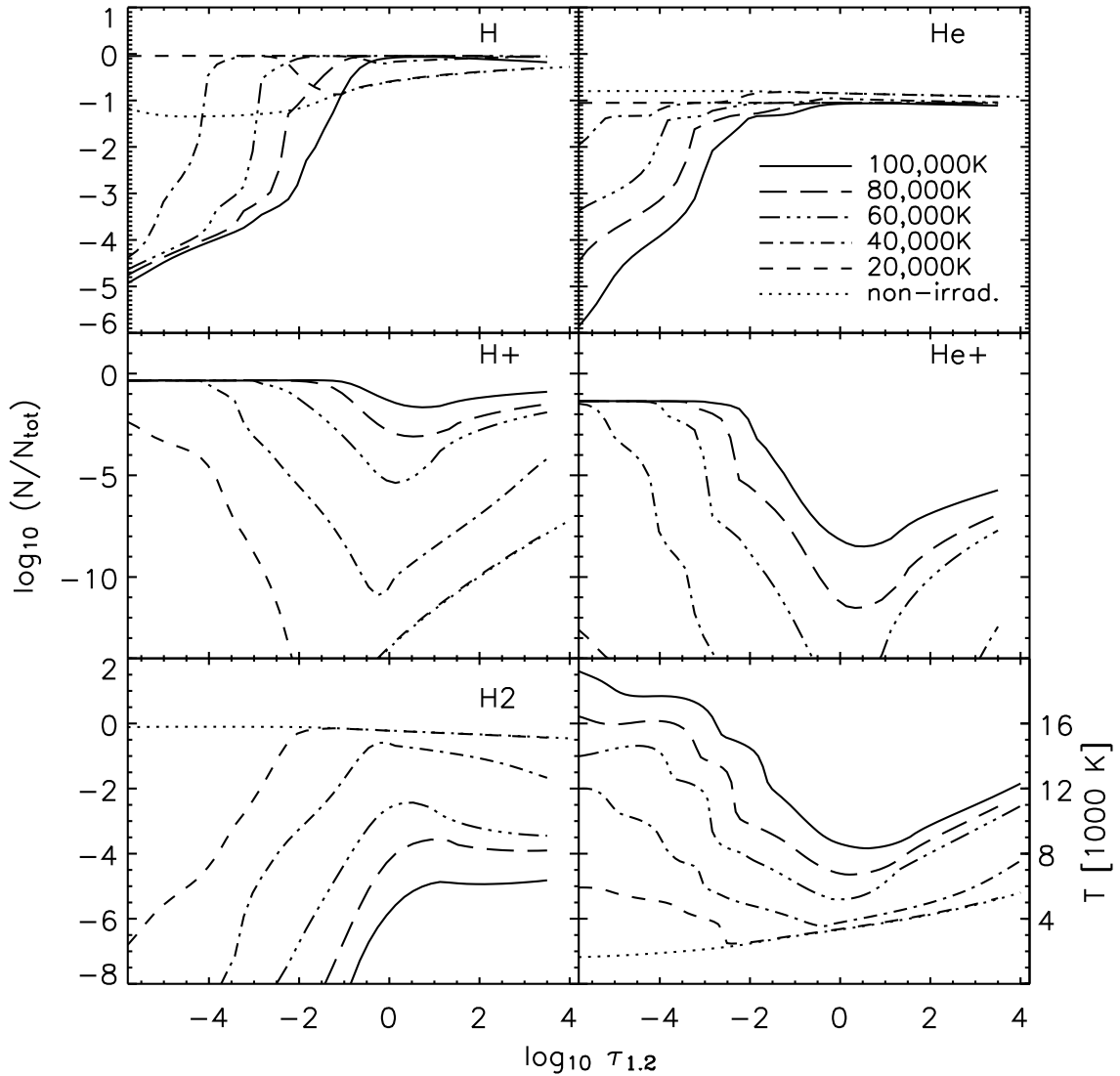}
\caption{
The concentrations (number density $N$ divided by the total number density
$N_{\rm tot}$) of H, H$^+$, H$_2$, He, and He$^+$ are shown for a $T_{\rm eff} =
3000$K non-irradiated model and the irradiated models with $T_{\rm p}$  =
20,000K, 40,000K, 60,000K, 80,000K, and 100,000K.  The corresponding
temperature structures are shown in the bottom right panel.
\label{numden1}}
\end{figure*}

\section{Results}
\subsection{Comparison to the BS93 Results}

Figure \ref{brett1} compares the temperature-pressure (T-P) profiles of
non-irradiated and irradiated models to similar calculations from BS93 (see
their Fig. 3).  The models have $T_{\rm int} = 3400\rm K$, $\log(g) = 5.0$ (cgs
units) and \Ms\ = 0.22 \msolar\ (note, BS93 used plane parallel models and so
\Ms\ is irrelevant).  The \phx\ irradiated model included a primary located at
an orbital separation of 0.39 $\rsolar$ with $T_{\rm p}$  = 10,000 K, \Rp =
0.0319$\rsolar$, and $\alpha = 1.0$.  These parameters were chosen to match the
incident flux used by BS93\nocite{Brett93}.

The models in Fig. \ref{brett1} illustrate the basic properties of an
irradiated atmosphere.  As should be expected, substantial heating occurs in
the upper layers of the irradiated atmosphere (low $P_{\rm gas}$) where
temperatures exceed those in the non-irradiated case by 3000K or more.  The
temperature also increases throughout the atmosphere, even at regions with high
gas pressures, resulting in a different inner adiabat compared to the
non-irradiated model (the reasons for this are discussed below).  Also, the
radiative-convective boundary nearly reaches $P_{\rm gas} \sim 10^6$ dynes
cm$^{-2}$ as convection retreats to deeper layers in the irradiated atmosphere.

Both \phx\ and BS93 models predict a similar temperature inversion at the top
of the irradiated atmosphere and similar adiabats at high $P_{\rm gas}$.
However, the BS93\nocite{Brett93} profiles are noticeably cooler than the \phx\
profiles at nearly all $P_{\rm gas}$.  Above the temperature minimum in the
irradiated case (near $P_{\rm gas} = 10^5$ dynes cm$^{-2}$), the temperatures
are more than 800K higher in the \phx\ models.  These differences are likely
due to the use of ``straight mean'' (SM) molecular opacities and a lack of any
atomic line opacity in the BS93 models.  The lack of atomic line opacity can
explain the differences at high pressures due to an underestimated backwarming
affect.  At low pressures, it is difficult to pinpoint the exact causes of the
differences but, as pointed out by BS93, SM opacities tended to overblanket
their models which could have caused preferential heating of the outer layers.
Note that, when the BS93 models were produced, reliable atomic and molecular
line-lists were only just becoming available and thus the use of SM opacities
was quite common.  Large line databases, however, have been steadily produced
over the past few years and are incorporated into \phx\ using a direct opacity
sampling method.  A second major difference between the two sets of models is
the completeness of the equation of state. \phx\ includes a far greater number
of species (literally many hundreds of molecules) in the solution of the
chemical equilibrium equations compared to the BS93 models.  Differences in the
chemical equilibrium can lead to very different concentrations of primary
opacity sources and ultimately very different temperature structures.  Despite
the differences between the models, the fact that two independent atmosphere
codes produce qualitatively consistent results is encouraging.

\begin{figure}
\plotone{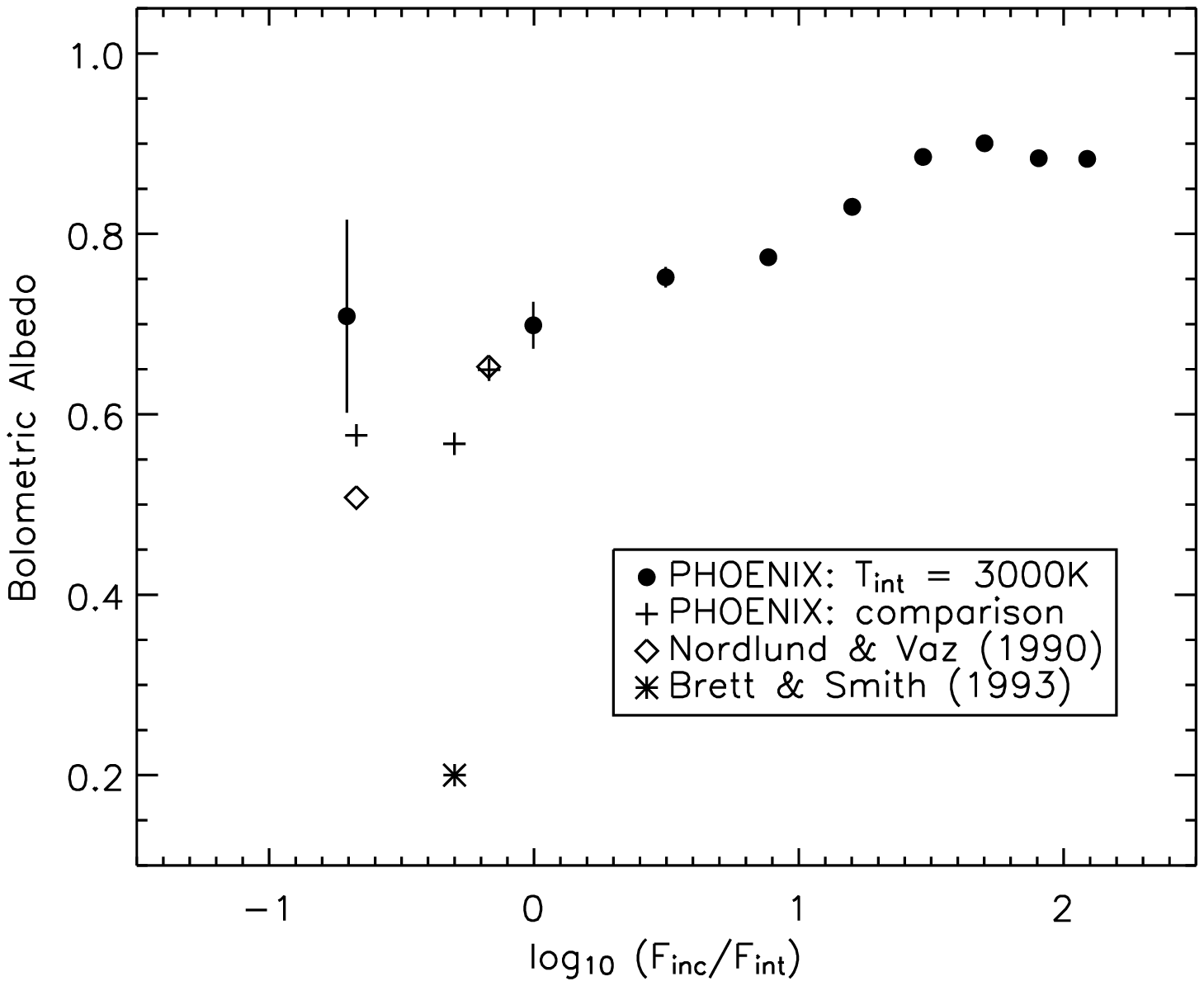}
\caption{
The bolometric albedo for the irradiated atmospheres in Fig. \ref{strucs1}
(filled symbols).  Error bars that assume a 2\% error in the net flux of the
model are given for each $T_{\rm int} = 3000$K model.  For large $F_{\rm inc}$,
the error bars are smaller than the symbol size.  Also plotted are albedos from
\cite{NordlundVaz1990} (diamonds) and BS93 (asterisk) along with three \phx\
models (pluses) specifically for comparing with these earlier works. 
\label{albedo}}
\end{figure}

\begin{figure*}
\plotone{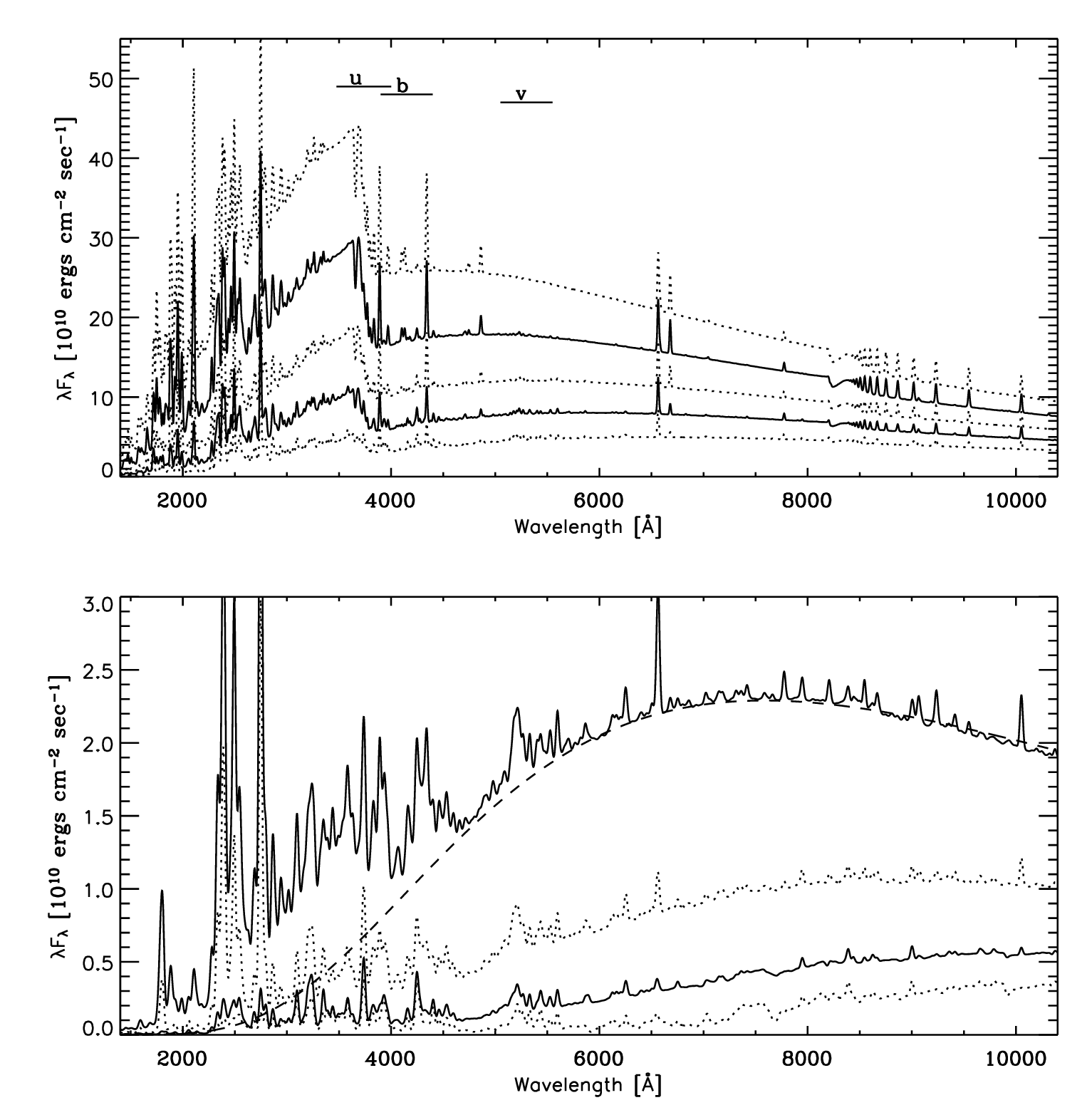}
\caption{
Spectra for the irradiated models shown in Fig. \ref{strucs1}.  $T_{\rm p} = $
20,000K to 50,000K in the lower panel and $T_{\rm p} = $ 60,000K to 100,000K in
the upper panel (with $T_{\rm p}$ increase from top to bottom).  To facilitate
comparison, the resolution was reduced to 15$\ang$ and alternating line styles
were used.  The dashed line is a best fitting black body spectrum (with $T_{\rm
eff} = 4840K$) for the $T_{\rm p} =$ 50,000K model..
\label{specs1}}
\end{figure*}

\subsection{Atmospheric Structures}
To explore the effects of irradiation by primaries with different $T_{\rm p}$
values, a sequence of models was calculated for a secondary with $T_{\rm int} =
3000\rm K$, $\log(g) = 5.0$, \Ms = 0.22 \msolar\ and solar metalicities.  The
primary was a $0.014\rsolar$, $\log(g)=8.0$ (sub-solar metalicity) WD located
$1 \rsolar$ from the secondary.  $T_{\rm p}$  was increased from 20,000K to
100,000K by steps of 10,000K.  With this setup, the incident flux spans almost
3 orders of magnitude with $-0.7 \le \log(F_{\rm inc}/F_{\rm int}) \le 2.1$.

The temperatures as functions of optical depth ($\taustd$) are shown in Fig.
\ref{strucs1} for each irradiated model atmosphere.  The models are fully
converged (i.e., energy is conserved to better than 2\% at all layers) and the
atmospheric chemistry and opacities are consistent with the changes in
temperature.  For the lowest primary temperature ($T_{\rm p} = 20,000\rm K$), a
significant temperature inversion forms in the upper regions of the atmosphere
while the deeper layers (near $\taustd = 1$) remain fairly similar to a
non-irradiated atmosphere with $T_{\rm eff} = T_{\rm int}$.  As $T_{\rm p}$
increases, the temperature inversion also increases and the temperature minimum
moves to deeper layers.  With $T_{\rm p}$  = 100,000K, the temperatures at the
top of the atmosphere reach 19,000K and the T-minimum is located at $\taustd <
1.0$.  Once $T_{\rm p}$  is higher than 30,000K, the inner temperature profile
is no longer similar to that of a standard M dwarf, despite the fact that
$T_{\rm int}$ was held fixed for this sequence of models.

The large range of atmospheric temperatures and pressures displayed in Fig.
\ref{strucs1} is accompanied by a very diverse chemical make-up.  In
non-irradiated M dwarfs, molecules (H$_2$, H$_2$O, CO, TiO, etc.) are the major
sources of opacity throughout the atmosphere.  For an irradiated atmosphere
with $T_{\rm p}$  = 20,000K, the incident flux raises the temperatures in the
upper atmosphere so much that most molecules cannot survive.  Figure
\ref{numden1} illustrates the changes to the number densities of H$_2$, H,
H$^+$, He, and He$^+$.  When $T_{\rm p}$  = 20,000K, the total concentration of
molecular species (N$_{\rm mols}$/N$_{\rm gas}$) at $\taustd \le 10^{-2}$
decreases by more than 5 orders of magnitude compared to the non-irradiated
case.  Also, a significant fraction of hydrogen is ionized near the top of the
atmosphere.  When $T_{\rm p} > 40,000$K, molecules are no longer important in
the photosphere and the concentration of H$_2$ at the temperature minimum is
less than $10^{-2}$.  Other molecules (e.g. CH$_4$ and CO) follow a similar
pattern as H$_2$ but with smaller concentrations at all layers. At even larger
$T_{\rm p}$, the atmosphere becomes mostly composed of atoms and ions and, when
$T_{\rm p} = 100,000\rm K$, hydrogen is almost completely ionized down to
$\taustd = 0.1$ and He is ionized down to $\taustd = 10^{-2}$.  The ionization
in the upper half of the atmosphere dramatically increases the number density
of free electrons which, in turn, causes the number density of H$^+$ and He$^+$
to increase.  The heating of the inner atmospheric layers also leads to
increased electron densities and concentrations of H$^+$ and He$^+$.  Overall,
the number densities of electrons, H$^+$, and He$^+$ follow the same pattern as
the temperature structure.  Note that, while the solution of the EOS in \phx\ allows for
the possible formation of dust grains and other condensates (as described in
Allard et al. 2001\nocite{AMES-2001}), all of the models presented here have
atmospheric temperatures that are too hot for grains to form.  For
intrinsically cooler secondaries (e.g., brown dwarfs and extrasolar planets)
and relatively low irradiation, grain formation will be very important
\cite[]{Barman01}.
 
\subsection{Entropy Matching}
The increase in temperature at large depths (i.e., large $P_{\rm gas}$) and
throughout the convection zone has been discussed in previous papers (Vaz \&
Nordlund 1985\nocite{VazNordlund1985}; BS93\nocite{Brett93}).  As described in
section \ref{modcon}, the model atmospheres are constrained to be in thermal
equilibrium and, therefore, each layer in the convection zone was required to
transport a constant total energy.  Furthermore, in a purely adiabatic
convection zone, the energy transport is determined by the adiabatic
temperature gradient.  Thus, a temperature increase near the top of the
convection zone must lead to an increase at all other layers in the convection
zone to ensure a constant luminosity is maintained at all layers of the
atmosphere.  The now well known consequence of these constraints is that
convective irradiated and non-irradiated models with the {\em same} effective
temperature (i.e. $T_{\rm int} = T_{\rm eff}$) do not reach the same adiabat at
depth (as illustrated in Figs.  \ref{brett1} and \ref{strucs1}).  Furthermore,
since the material below the photosphere is likely to be well mixed, the
irradiated and non-irradiated model atmospheres should have the same entropy at
depth if they are to represent the two different hemispheres of the same object
(Vaz \& Nordlund 1985 \nocite{VazNordlund1985}; BS93\nocite{Brett93}).  In most
cases, entropy matching can be achieved by lowering $T_{\rm int}$ of the
irradiated model until the T-P profile matches the non-irradiated model in the
convection zone.  The difference between $T_{\rm int}$ and $T_{\rm eff}$ for
the matching irradiated and non-irradiated atmospheres is a measure of the
bolometric reflection albedo and the energy that is horizontally redistributed
in the photosphere (BS93)\nocite{Brett93}.   

The entropy matching technique is designed to mimic the anticipated conditions
at large optical depths within the convection zone.  These layers should
undergo efficient mixing over very short timescales ensuring a constant
internal entropy across both hemispheres.  Also, the decrease in the
atmospheric temperature gradient on the heated hemisphere will slow the rate at
which internal energy can escape from that side of the secondary implying a
smaller {\em intrinsic} luminosity on the heated hemisphere compared to the
non-heated side.  This is consistent with the fact that entropy matching models
will always have intrinsic effective temperatures that are less than (or equal
to) the $T_{\rm eff}$ on the non-irradiated side.  Entropy matching, however,
cannot deal with the potential horizontal flow of energy in the radiative
zones.

\begin{figure*}
\plottwo{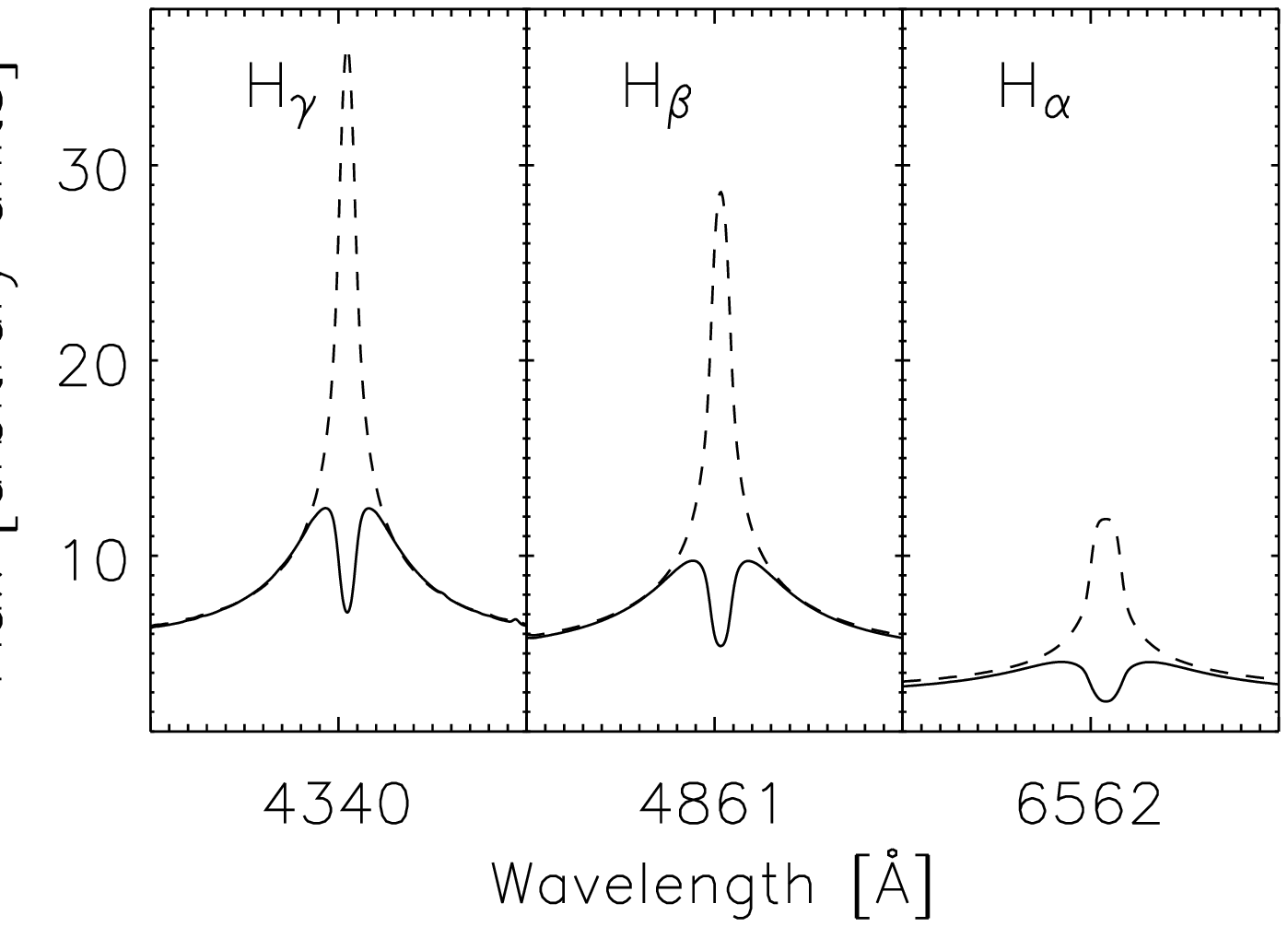}{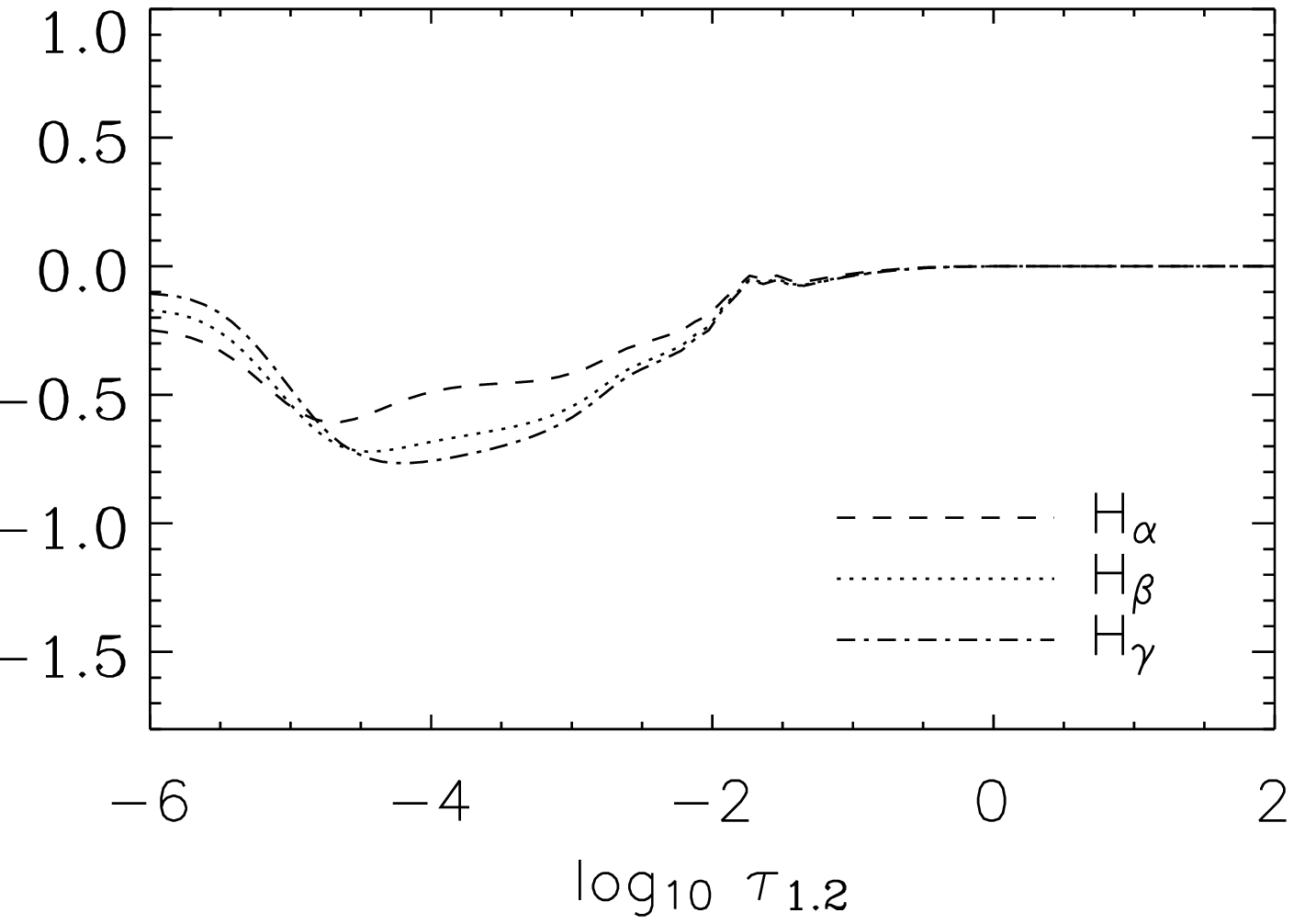}
\caption{
Right panel: the ratio of departure coefficients for the levels involving
H$_\alpha$, H$_\beta$, and H$_\gamma$ as functions of optical depth at
1.2\micron.  Each ratio is proportional to the non-LTE line source function. A
value other than 1 indicates departure from LTE.  Left panel: comparison
between LTE (dashed) and non-LTE (solid) line profiles for the corresponding
Balmer lines. Each panel is $20\ang$ wide.
\label{nlte1}}
\end{figure*}

To find pairs of entropy matching models, a sequence of non-irradiated models
was computed with 2800K $< T_{\rm eff} <$ 6000K.  Then, using the entropy at
the bottom of an irradiated model, the matching effective temperature ($T_{\rm
matched}$) was found by linearly interpolating along a $S(T_{\rm eff})$ curve
at the same $P_{\rm gas}$ in the non-irradiated models.  This method, which is
very similar to that described by \cite{VazNordlund1985}, was applied to each
of the models shown in Fig. \ref{strucs1}.  The matched, non-irradiated,
atmospheric structures for the $T_{\rm p}$  = 20,000K, 50,000K, and 100,000K
models are also shown as dashed lines and have, respectively, $T_{\rm eff} =
3041\rm{K}, 3857\rm{K}$, and $5910\rm{K}$.  Recall that the corresponding
irradiated atmosphere models all have $T_{\rm int} = 3000\rm K$.  

The two effective temperature values for the pair of entropy matching models
can be used to obtain a value for the bolometric reflection albedo ($w$) via
the expression,
\begin{equation}\label{weq}
(1 - w) F_{\rm inc} = \sigma (T_{\rm matched}^4 - T_{\rm eff}^4) 
\end{equation}
(BS93\nocite{Brett93}).  The albedos versus the ratio of incident to intrinsic
fluxes are shown in Fig.  \ref{albedo}.  For very large incident flux the
albedo approaches 1.  This is expected since the atmospheres become radiative
down to large depths as $F_{\rm inc}$ increases. From Fig. \ref{strucs1}, the
radiative convective boundary is at $\taustd \ll 1$ when $\log F_{\rm
inc}/F_{\rm int} > 0.5$.  Atmospheres that are completely in radiative
equilibrium have an albedo exactly equal to 1.0 \cite[]{Eddington}.  For low
incident flux the albedo decreases but is still well above the canonical value
of 0.5 for convective atmospheres.  In general, the \phx\ albedos are
consistent with the recently published albedos for a variety of models with
$T_{\rm int} \ge 4000{\rm K}$ and $\log F_{\rm inc}/F_{\rm int} \le 1$
\cite[Fig. 4]{Claret2001}.

All models in this paper were considered converged when the flux was conserved
to 2\% or better at all layers.  At this level of convergence, no appreciable
improvements in the models are obtained by further iterations.  None the less,
a small error exists in the net flux of the models which translates into a
small error associated with $T_{\rm matched}^4$ in eq.  \ref{weq}.  For small
values of $F_{\rm inc}$, the difference between $T_{\rm matched}$ and $T_{\rm
eff}$ is also small ($\sim 50$K) which means that even a 2\% error in the flux
can result in a large error in $w$.  Error bars corresponding to a 2\% error in
the model's net flux are shown for each albedo ($T_{\rm int} = 3000\rm K$
models) in Fig.  \ref{albedo}.  Since the error in $w$ is also proportional to
$1/F_{\rm inc}$, the error bars decrease with increasing $F_{\rm inc}$ (and the
difference between $T_{\rm matched}$ and $T_{\rm eff}$ is much larger).

Comparison \phx\ models were also constructed to match the input parameters of
several models from \cite{NordlundVaz1990} (see their Table 2; non-grey models
with mixing length parameter equal to 2) in addition to the
BS93\nocite{Brett93} models shown in Fig. \ref{brett1}.  The \phx\ abledos for
these models are plotted in Fig. \ref{albedo} and agree well with
\cite{NordlundVaz1990} while the BS93 albedo is less than half the value for
any of the others.  It is difficult to determine the exact reasons why the BS93
result is so much lower than the \phx\ albedo.  In order to obtain an albedo of
0.2 for the \phx\ model, $T_{\rm matched}$ would need to be $\sim 3800$K,
instead of the current value of $\sim 3600\rm{K}$.  Note that the BS93 models
suffered from many competing simplifications that included SM opacities, the
absence of atomic line opacity, and the use of underestimated astrophysical
values for the TiO oscillator strengths.  The impact of these simplifications
can be seen by comparing the BS93 non-irradiated models to Brett's later, much
improved, non-irradiated M dwarf models \cite[]{Brett1995}.  The temperature at
the bottom of a $T_{\rm eff} = 3800$K model from \cite{Brett1995} is nearly
1000K hotter than the bottom of the non-irradiated model shown in BS93 with the
same $T_{\rm eff}$ (their Fig. 7).  However, at $T_{\rm eff} = 3400$K, the
differences at the bottom are on the order of a few 100K.  Using the improved
\cite{Brett1995} non-irradiated models to find the $T_{\rm matched}$ for the
BS93 irradiated model leads to an albedo very comparable to the \phx\ value.
The good agreement in Fig.  \ref{brett1} between the \phx\ and BS93 irradiated
models suggests that, for this particular set of parameters, the extrinsic
heating offsets many of the opacity limitations in the BS93 models.

\subsection{Synthetic Spectra}\label{spectra_section}

The changes to the temperature structure and chemical composition have dramatic
consequences for the spectra emerging from the illuminated hemisphere.  The
synthetic spectra for each of the structures in Fig.  \ref{strucs1} are
displayed in Fig. \ref{specs1}.  The atmospheric model irradiated by the
coolest WD ($T_{\rm p} = 20,000\rm K$) retains many of the infrared spectral
features common to M dwarfs, e.g. TiO and VO absorption bands.  However, a
substantial amount of flux emerges at short wavelengths with a dense spectrum
of narrow emission lines.  As $T_{\rm p}$ increases, the flux emerges from two
fairly distinct spectral regions divided by a pronounced Balmer-edge at
3647\AA.  When $T_{\rm p}$ reaches $30,000\rm K$, the broad spectral features
in the red portion of the secondary's spectrum disappear and are replaced by a
few narrow emission lines.  The blue-UV portions of the spectra, however, have
complex mixtures of emission and absorption lines.  The strongest of these
lines are from \ion{Fe}{1}, \ion{Fe}{2}, \ion{Mg}{1}, \ion{Mg}{2}, and
\ion{Ni}{1}.  When $T_{\rm p} = 100,000$K, many additional emission lines from
\ion{Fe}{3}, \ion{Al}{2}, \ion{Ti}{2}, \ion{Ni}{2}, and \ion{Si}{1} appear in
the UV--blue spectrum.

This division of the spectrum into two distinct components is a consequence of
the two very different temperature regimes in the atmosphere.  The hot,
inverted, outer temperature structure produces the blue-UV portion of the
spectrum. Note, $\tau = 1$ at 3000\AA\ is at roughly $\taustd = 10^{-4}$.  The
red part of the spectrum forms near $\taustd = 1$, while the core of most
emission lines form at smaller $\taustd$.  For all of the cases presented in
Fig. \ref{specs1}, the incident radiation is sufficient to produce Balmer, He,
and metal emission lines.  

A best fitting black body SED, with $T_{\rm eff} = 4840\rm K$, is also shown in
Fig. \ref{specs1} for $\lambda \ge 4500\ang$ and $T_{\rm p} = 50,000\rm K$.
Apart from the emission lines, the red portions of the spectra resemble black
body SEDs when $T_{\rm p} > 30,000\rm K$.  This is mostly due to the roughly
isothermal $T-P$ structure near the temperature minimum that occurs close to
continuum forming region ($\taustd = 1.0$).  Spectral features diminish where
the temperature structure is flat because the LTE source function cannot vary
with depth.  At shorter wavelengths, the best fitting black body underestimates
the emergent flux by orders of magnitude.   Also, for $T_{\rm p} \ge 60,000\rm
K$, a steep Blamer-edge forms in emission.  A steep Blamer-edge could explain
why, for certain binaries, it is difficult to obtain a single effective
temperature for the secondary when using black body SEDs to fit $ubv$
photometry.

\begin{deluxetable*}{ccccccccccl}
\tabletypesize{\footnotesize}
\tablecolumns{11}
\tablewidth{0pc}
\tablecaption{System properties for example pre-CVs}
\tablehead{
\colhead{}  & \multicolumn{3}{c}{Primary} & \colhead{} & \multicolumn{3}{c}{Secondary}
 & \colhead{} & \colhead{} & \colhead{} \\
\cline{2-4} \cline{6-8} \\
\colhead{Object} & \colhead{T$_{\rm eff}$(K)} & \colhead{R(R$_\odot$)} &
\colhead{M(M$_\odot$)} & \colhead{} &\colhead{T$_{\rm eff}$(K)} &
\colhead{R(R$_\odot$)} & \colhead{M(M$_\odot$)} &
\colhead{a(R$_\odot$)} & \colhead{$\log(F_{\rm inc})$} & \colhead{refs.}}
\startdata
GD 245 & 22,170 & 0.015&0.48 && 3560 & 0.27& 0.22 & 1.17 &  9.35 &\cite{schmidt95}\\
NN Ser & 55,000 & 0.019&0.57 && 2900 & 0.17& 0.12 & 0.95 & 11.32 &\cite{catalan94}\\
AA Dor & 42,000 & 0.19 &0.40 && 3000 & 0.10& 0.073& 1.14 & 12.69 &\cite{Hilditch2003}\\
UU Sge & 85,000 & 0.34 & 0.63&& 6250 & 0.54& 0.29 & 2.46 & 13.79 &\cite{Bell1994}\\
\enddata
\label{examples}
\end{deluxetable*}

\subsection{Hydrogen in non-LTE}
A striking feature of the spectra shown in Fig. \ref{specs1} are the hydrogen
emission lines. However, one caveat should be mentioned for the model results
discussed so far; in all of the above cases LTE was assumed.  Given the {\em
non-local} origin of a primary source of energy (the incident flux), the
conditions in the upper atmosphere, where gas pressures are low and
temperatures are high, are well suited for significant departures form LTE.  To
test the importance of non-LTE effects,  the $T_{\rm p}$  = 100,000K model was
recalculated with neutral Hydrogen allowed to depart from LTE.  For this
particular test, the Hydrogen model atom included 31 levels, and 435 primary
transitions (all bound-bound transitions with $\log(gf) > -2.0$) were included
in the solution of the statistical equilibrium equations.  For details of the
numerical methods used to solve the rate equations, see \cite{jcam}.  

The LTE and non-LTE line profiles for H$_\alpha$, H$_\beta$, and H$_\gamma$ are
compared in Figure \ref{nlte1}.  Both LTE and non-LTE cases have Balmer wings
in emission while dramatic differences exist between the line cores.  The
non-LTE line profiles all have reversed cores while the LTE lines have cores
that are substantially brighter than the non-LTE case.  The non-LTE H$_\alpha$
line core drops slightly below the continuum flux level while the cores of
H$_\beta$, and H$_\gamma$ are only slightly above the continuum.  The ratio the
LTE and non-LTE Balmer line fluxes, for this case, are not significantly
different.  The non-LTE flux ratio H$_\alpha$/H$_\beta$ is $\sim 0.52$
compared to $\sim 0.57$ for the LTE case.   The non-LTE flux ratios for
H$_\alpha$/H$_\gamma$ and H$_\beta$/H$_\gamma$ are $\sim 0.41$ and $\sim 0.80$,
respectively.

In both non-LTE and LTE cases, the line wings form near $\taustd = 1$ while the
core of the lines form much higher in the atmosphere near $\taustd = 10^{-4}$.
In LTE, the depth dependent line source function is simply a black body with
temperature equal to the local gas temperature.  Consequently, the LTE line
profile mimics the temperature inversion (top curve in Fig. \ref{strucs1})
which changes by $\sim 9000$K between these two optical depths.  However, when
the LTE approximation is removed, noticeable departures from the LTE atomic
level populations occur for the $n=2$ level, which is over populated by a
factor 10 (compared to the LTE populations) at $\taustd = 10^{-4}$.  The $n=1$
and $3 < n < 6$ levels are also slightly over populated in the upper
atmosphere, but less so compared to $n=2$.  The non-LTE line source function is
not a Planck function, but instead is related to the ratio of the departure
coefficients ($b_i$'s) for the two levels involved in the transition.  The
ratio of the $b_i$'s for these three Balmer lines are shown in Fig.
\ref{nlte1}.  For all three Balmer lines the ratio of the $b_i$'s, and
consequently the line source function, decreases in the very region where the
line cores form resulting in a fainter core compared to the wings.  The LTE and
non-LTE profiles are similar in the wings since the $b_i$'s approach unity at
large optical depth.  It is likely that other lines are also affected by
non-LTE affects and the authors are working on a future paper that fully
explores these issues for a number of atoms and ions.

\subsection{pre-CV Examples}
The models and methods described above were applied to four example systems (GD
245, NN Ser, AA Dor, and UU Sge) that have been well observed and documented in
the literature.  These systems were chosen because they represent the broadest
range of properties for PCEBs, spanning 4 orders of magnitude in incident flux.
Two systems, GD 245 and NN Ser, have WD primaries while the other two, AA Dor
and UU Sge, have hot sub-dwarf stars as primaries.  The purpose of this section
is not to determine the properties of each system, but instead to illustrate
some of the probable characteristics of their secondary's heated hemisphere.
The orbital parameters and the bulk properties for the primary and secondary
were taken from previous works and are listed in Table \ref{examples}.

\subsubsection{GD 245 \& NN Ser}
GD 245 and NN Ser are two examples of pre-CVs for which phase resolved
spectroscopy has been performed \cite[]{schmidt95,catalan94}.  Both systems are
observed to have Balmer lines composed of absorption wings from the primary and
narrow emission cores from the heated face of the secondary.  The strength of
the emission peaks at zero phase when the heated hemisphere is most visible.
The observations of GD 245 clearly show the presence of additional emission
lines from the secondary due to various metals (\ion{Mg}{2}, \ion{Fe}{1},
\ion{Na}{1}, and \ion{Ca}{2}).  The spectra of NN Ser also show the presence
of emission from \ion{Ca}{2} and \ion{He}{1}.  Using the parameters listed in
Table \ref{examples}, model atmospheres were computed for the secondary of
both GD 245 and NN Ser. As described above, a single irradiated model was used
to represent the average properties of the heated hemisphere.  

\begin{figure}
\plotone{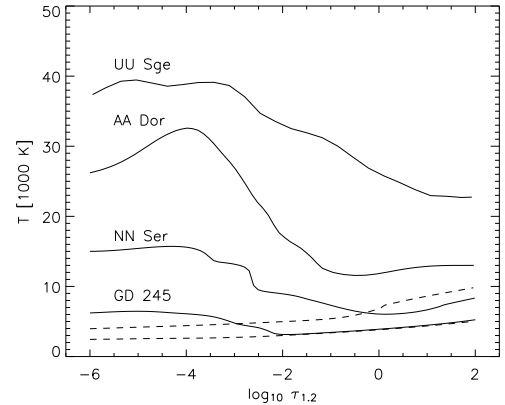}
\caption{
Temperature versus optical depth (at 1.2\micron) from irradiated models for the
example cases listed in Table \ref{examples} (solid lines).  Dashed lines are
non-irradiated models with T$_{\rm eff} = 6000K$ (top) and T$_{\rm eff} = 3000K$
(bottom). 
\label{examples_strucs}}
\end{figure}

The temperature structures are shown in Fig. \ref{examples_strucs}.  Both GD
245 and NN Ser have substantial temperature inversions at the top of the
atmosphere and, in the case of NN Ser, extending down to $\taustd = 1$.  Both
structures are compared to that of a non-irradiated M dwarf with $\log(g) = 5.0$
and $T_{\rm eff} = 3500\rm K$ (lowest dashed curve in Fig.  \ref{examples_strucs}).
Apart from the temperature inversion, the irradiated model of GD 245 remains
similar  to a normal M dwarf atmosphere with convection present at layers where
$\taustd < 0.1$.  However, the model for NN Ser, which included a much hotter
primary ($T_{\rm p}$  = 55,000K) than GD 245, has a purely radiative atmosphere and
is nearly isothermal down to $\taustd = 1$. 


\begin{figure*}
\plotone{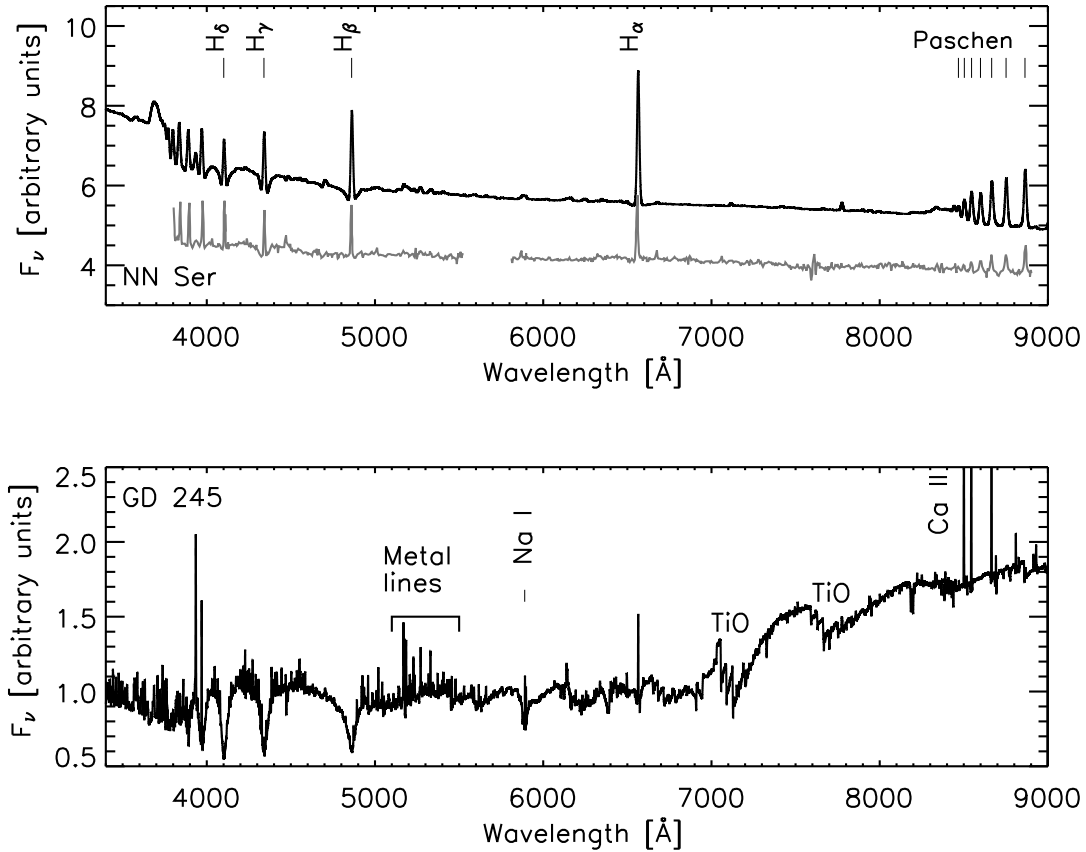}
\caption{
Combined spectra, i.e. primary flux plus the flux from the irradiated
hemisphere of the secondary, for NN Ser (top) and GD 245 (bottom).  The
atmospheric structures for the irradiated model atmospheres are shown in Fig.
\ref{examples_strucs}.  The primary and secondary spectra were combined
assuming an orbital inclination of 90$^\circ$ and phase 0.  Note, both figures
were designed to mimic those shown in \cite{schmidt95} and \cite{catalan94} to
facilitate a comparison. Furthermore, both secondary spectra were rotationally
broadened using $V_{\rm rot} = 70$ km sec$^{-1}$ and the resolution was reduced
to $5\ang$ for NN Ser and $2\ang$ for GD 245 to matched the observed spectral
resolutions from these studies.  In the top panel, an observed spectrum 
of NN Ser from \cite{catalan94} is also shown, but offset to make the
plot more readable.  Note, that the meanings of phases 0 and 0.5 are reversed
in the \cite{catalan94} paper compared the present work. 
\label{gd245_NNser_specs}}
\end{figure*}

In the lower panel of Fig. \ref{gd245_NNser_specs}, a {\em total} synthetic
spectrum is shown for the GD 245 system: i.e. the spectrum of the WD primary
and heated face of the secondary were added using the parameters listed in
Table \ref{examples} assuming a 90$^\circ$ inclination.  At $\lambda \simle
6000\ang$, the pseudo-continuum is shaped primarily by the WD's continuum and
broad Balmer absorption lines.  The red part of the spectrum is dominated by
the flux from the secondary, which retains deep molecular absorption bands
despite the extrinsic heating.  In addition, the synthetic spectrum
approximately reproduced the basic collection of emission lines and relative
line strengths shown in previously published observed spectra (see Figs. 2 and
3 of Schmidt et al. 1995\nocite{schmidt95}).  There are certainly differences
between the observed spectrum and the synthetic spectrum (e.g., the number and
widths of the metal lines).  However, given that only a single 1-D model was
used for the secondary's heated hemisphere and only Hydrogen was treated in
non-LTE, the similarities are encouraging.  Also, \cite{schmidt95} estimated
that GD 245 emits about 60 times more H$_\alpha$ flux than predicted by simply
counting the raw number of incident ionizing photons from the WD.
\cite{schmidt95} suggested that irradiation might reduce the number densities
of molecular and neutral atomic opacity sources and allow for the WD's Balmer
continuum to supply the photons needed to explain the H$_\alpha$ flux.  This
idea is supported by the current models, which do show a dramatic decrease in
the concentration of molecules and neutral atomic species.  Furthermore, the
synthetic spectrum shown in Fig. \ref{gd245_NNser_specs} has H$_\alpha$ flux
$\sim 2\times 10^{-14}$ ergs cm$^{-2}$ sec$^{-1}$ which is close to the
observed value of $3.7 \times 10^{-14}$ ergs cm$^{-2}$ sec$^{-1}$ at phase
zero.  Note that a more realistic synthetic spectrum built from many irradiated
models that simulate the variable heating across the secondary's surface
reproduces the observed Balmer line profiles of GD 245 very well
\cite[]{myphd}.

The temperature structure of the heated hemisphere in NN Ser is significantly
different from that of any non-irradiated main sequence star. The hot WD flux
heated the secondary enough to suppress convection below the photosphere and
reduced the presence of most molecules to very low concentrations.  The
combined WD + M dwarf synthetic spectrum for NN Ser is shown in Fig.
\ref{gd245_NNser_specs} (top panel) assuming a 90$^\circ$ inclination.  As with
GD 245, the flux from the hot primary determined most of the pseudo-continuum
while narrow Balmer and Paschen lines filled in the cores of the WD hydrogen
absorption lines.  Also shown in Fig. \ref{gd245_NNser_specs} is a digitized
version of the observed spectrum of NN Ser published by \cite{catalan94} for a
similar phase.  As with GD 245, the synthetic spectrum is very comparable to
the observed spectrum of NN Ser.  The pseudo-continuum and most emission lines
are well reproduced by the model.  The most noticeable differences are the
strengths of the H emission lines which are weaker in the red half of the
observed data.  It should be stressed that no attempt has been made yet to find
the best-fitting model for the observed spectrum.  This comparison is simply
for illustrative purposes and a more detailed analysis of observed spectra will
be the subject of a later paper.

\subsubsection{AA Dor \& UU Sge}

\begin{figure}[t]
\plotone{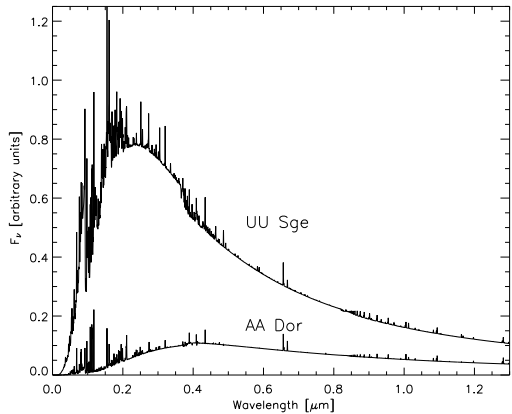}
\caption{
Spectra for the irradiated hemispheres of AA Dor and UU Sge.  The atmospheric
structures for the irradiated model atmospheres are shown in Fig.
\ref{examples_strucs}. For both spectra, the resolution was reduced to $5\ang$.
\label{AADor_UUSge_specs}}
\end{figure}

AA Dor and UU Sge are examples of extreme irradiation. The primaries, in both
cases, are hot sub-dwarfs with radii more than an order of magnitude larger
than the WD primaries of GD 245 and NN Ser.  Furthermore, the primary of UU Sge
is extremely hot, $T_{\rm p} $  = 85,000K.  The atmospheric structures for the
heated hemispheres of UU Sge and AA Dor are shown in Fig.
\ref{examples_strucs}.  Both models have large temperature inversions and are
radiative to depths well below the photosphere.  Temperatures in AA Dor rise by
nearly 20,000K between $\taustd = 1$ and $\taustd = 10^{-4}$.  In the heated
hemisphere model for UU Sge, temperatures are well above 20,000K with a maximum
temperature near $\taustd = 10^{-4}$ of almost 40,000K.  The impact of this
heating is also evident in their spectra.  Figure \ref{AADor_UUSge_specs} shows
the flux from the heated hemisphere model only; for these systems the dominant
flux is from the primary and the secondary simply makes a small contribution to
the overall pseudo-continuum.  As in Fig. \ref{specs1}, a number of emission
lines form in the bluest parts of the spectrum.

Based on light curve analyses, it has been estimated that the temperature varies
from 20,000K and 2000K across the heated and non-heated faces of AA Dor
\cite[]{Hilditch2003}.  Such a difference is consistent with the irradiated
and non-irradiated structures from Fig.  \ref{examples_strucs} (keeping in mind
that the models are averages for each hemisphere).  The light curve analyses
also suggest that the albedo is very close to one \cite[]{Hilditch2003}  which
is also consistent with the model being fully radiative (and Fig. \ref{albedo}).

\section{Discussion and Conclusions}

Model atmospheres have been computed for irradiated stars in pre-CV systems and
high-resolution synthetic spectra are now available for direct comparison to
observations.  The models presented above illustrate the dramatic differences
that may exist between the heated and non-heated hemispheres of the secondary
in pre-CV systems.  Large temperature inversions, which can extend down to the
continuum forming layers of the photosphere, are predicted on the heated
hemisphere.  Such temperature inversions naturally lead to emergent spectra that
are unlike normal main sequence stars and are very different from black bodies.
A steep Balmer-edge and strong hydrogen emission lines are present in addition
to a forest of narrow metal lines in emission.  It has also been shown that
departures from LTE are to be expected and can have significant impact on the
hydrogen line profiles.  The present models are consistent with previous
findings for the atmospheric structures and bolometric reflection albedos (Vaz
\& Nordlund 1985; Nordlund \& Vaz 1990; BS93)\nocite{VazNordlund1985,
NordlundVaz1990, Brett93}.

The large differences between the irradiated and non-irradiated atmosphere
models suggest that steep horizontal temperature and pressure gradients may
exist across the surface of a pre-CV secondary.  If present, the secondary's
heated hemisphere will have a SED that varies substantially across the heated
surface, which may be underrepresented by single 1-D averaged models.  Such
gradients could also lead to strong horizontal flows and redistribution
of energy to the non-irradiated hemisphere \cite[]{Kirbiyik1976,Beer2002}.  
Furthermore, these gradients may lead to convectively unstable layers
above the radiative-convective boundary predicted by the mixing length theory.
The present work will be expanded to include full surface modeling of pre-CV
secondaries in order to compute even more realistic light curves and synthetic
spectra.  

A detailed non-LTE analysis is currently underway for a large number of
atoms and ions for the strongly irradiated systems.  For the cooler cases that
still have significant concentrations of molecules in their atmosphere, solving
the rate equations is a very challenging problem.  Much of the atomic and
molecular data (e.g.  collisional cross-sections for atom/ion plus molecule
interactions) are not yet available and are difficult to obtain.  However, some
useful limits can be placed on the magnitude of the non-LTE effects for such
cases.

High-resolution, phase-resolved, spectra have been recently obtained for
several interesting pre-CVs \cite[]{Exter2003b,Exter2003a} and a future paper
will be devoted to an analysis of these systems using synthetic spectra from
irradiated model atmospheres.  The example pre-CV systems described above are
also prime candidates for detailed comparisons between observed spectra and
synthetic spectra.  In addition to providing insight into the atmospheric
conditions of irradiated M dwarfs, such comparisons will also provide valuable
tests for the modeling techniques that are currently being used in other areas
(e.g. White Dwarf -- Brown Dwarf pairs and close-in extrasolar planets).

\acknowledgments 
T. B. wishes to thank Katrina Exter and David Alexander for many useful
discussions during the course of this work.  This research was partially
supported by the CNRS and NASA LTSA grant NAG5-3435 to Wichita State
University. T. B. also acknowledges support by NASA through the American
Astronomical Society's small research grant program.  Some of the calculations
presented in this paper were performed on the IBM pSeries of the HLRN, the IBM
SP of the San Diego Supercomputer Center (SDSC) with support from the National
Science Foundation, the IBM SP of the NERSC with support from the DoE, and the
Wichita State University High Performance Computing Center with the support of
Kansas NSF Cooperative Agreement EPS-9874732.  We thank all these institutions
for a generous allocation of computer time.


\end{document}